\documentstyle[12pt]{article}
\newcommand{\sbody}[2]{{\textstyle\frac{#1}{#2}}}

\begin{document}
\begin{center}
\vfill
\large\bf{Gauge Model With Extended Field Transformations}\\
\large\bf{in Euclidean Space}
\end{center}
\vfill
\begin{center}
D.G.C. McKeon$^{1,2}$\\
T.N. Sherry$^{1,2}$\\
$^1$Department of Applied \vspace{-.1cm}Mathematics\\
University of Western \vspace{-.1cm}Ontario\\
London\vspace{-.1cm}\\
CANADA $\;\;$ N6A 5B7\\
$^{2}$Department of Mathematical \vspace{-.1cm}Physics\\
National University of \vspace{-.1cm}Ireland\\
Galway\vspace{-.1cm}\\
IRELAND
\end{center}
\vfill
email: $^1$TMLEAFS@APMATHS.UWO.CA\\
$^{2}$TOM.SHERRY@NUIGALWAY.IE\\
Tel: (519)679-2111, ext. 8789\\
Fax: (519)661-3523
\eject
\begin{flushleft}
{\Large\bf{Abstract}}
\end{flushleft}

An $SO(4)$ gauge invariant model with extended field transformations is examined in
four
dimensional Euclidean space.  The gauge field is $(A^\mu)^{\alpha\beta} = \sbody12\,
t^{\mu\nu\lambda} (M^{\nu\lambda})^{\alpha\beta}$ where $M^{\nu\lambda}$ are the $SO(4)$
generators in the fundamental representation.  The $SO(4)$ gauge indices also participate in the
Euclidean space $SO(4)$ transformations giving the extended field transformations.  We provide
the decomposition of the {\underline{reducible}} field $t^{\mu\nu\lambda}$ in terms of fields
{\underline{irreducible}} under $SO(4)$.  The $SO(4)$ gauge transformations for the irreducible
fields mix fields of different spin.  Reducible matter fields are introduced in the form of a Dirac
field in the fundamental representation of the gauge group and its decomposition in terms of
irreducible fields is also provided.  The approach is shown to be applicable also to $SO(5)$
gauge models in five dimensional Euclidean space.

\section{Introduction}

Pure Yang-Mills gauge models with extended field transformations (or extended gauge models)
were introduced some time ago [1].  In these models the gauge group indices also participate in
the space-time ``Lorentz'' transformations.  Two features of such models of note are the
expansion of the gauge field in terms of a number of fields which are ``Lorentz'' irreducible, and
the mixing of these fields in the gauge transformations of the model.

In this paper we examine the special case of an extended gauge model in four dimensional
Euclidean space ($4dE$) with gauge group $SO(4)$.  The ``Lorentz'' transformations in $4dE$
are then $SO(4)$ transformations. The $SO(4)$ gauge group indices participate in the $SO(4)$
space-time transformations in a well defined way.  This model has recently been examined [2]
and shown to have a number of interesting features.  We now examine the field content of the
model in terms of fields irreducible under the ``Lorentz'' $SO(4)$ space-time transformations. 
We also introduce matter fields coupled to the gauge fields: the matter fields being Dirac 
4-spinors in the fundamental representation of the $SO(4)$ gauge group. For the matter fields
also the gauge group indices participate in the $SO(4)$ space-time transformations. As a
consequence, the matter fields also are $SO(4)$-reducible.  Their decomposition in terms of
$SO(4)$-irreducible fields is straightforward.

Once the decomposition of the gauge and matter fields has been established, it is a relatively easy
exercise to deduce the form of the $SO(4)$ gauge transformations for the irreducible fields.  In
these gauge transformations we observe the mixing of the different $SO(4)$-irreducible
representations. Indeed, in terms of their spin content, we see mixing in the same transformation
of
\begin{enumerate}
\item[(i)] fields of spin $1$ and fields of spin $2$ and spin $1$, and
\item[(ii)] fields of spin $\sbody12$ and fields of spin $\sbody32$ and
$\sbody12$.
\end{enumerate}
This mixing of fields of different spin is not to be confused with supersymmetry.  The set of
fields which mix together under the gauge transformations are always either integer spin
multiplets, or half-odd-integer spin multiplets. This mixing of fields of different spin under a
symmetry transformation provides a second example (in addition to supersymmetry) of how the
strictures of the Coleman-Mandula theorem can be circumvented. 

In a recent paper [3] we have examined in detail the spinor representations of $SO(4)$ and the
associated four- and two-component spinor fields in $4dE$.  The spinor representations of
$SO(4)$ are quite different from those of $SO(3,1)$, the Lorentz group of space-time
transformations in four dimensional Minkowski space.  We include in the appendices a review
of the salient features of this analysis together with a list of relevant mathematical formulae
required in our discussion.

The remainder of this paper is organized as follows.  In Section 2 we introduce the extended
$SO(4)$ gauge model.  In Section 3 we expand the $SO(4)$-reducible gauge field in terms of
$SO(4)$-irreducible gauge fields. In Section 4 we carry out the corresponding expansion for the
reducible matter spinor fields.  In Section 5 we examine the gauge transformations in terms of
the $SO(4)$ irreducible fields.  We also examine the Lagrangian density in terms of the $SO(4)$
irreducible fields.   In Section 6 we discuss some of the general features of the model, and we
indicate how the analysis can be extended to higher dimensions.

\section{Extended $SO(4)$ Gauge Model}

The Lagrangian density for an extended pure Yang-Mills gauge model in $4dE$ has the usual
structure in terms of the gauge field $A_\mu$ and the associated field strength $F_{\mu\nu}$,
namely
$$L_{YM} = \sbody14 Tr F^{\mu\nu} F^{\mu\nu}\eqno(1)$$
where
$$F^{\mu\nu} = \partial^\mu A^\nu - \partial^\nu A^\mu + i[A^\mu, A^\nu].\eqno(2)$$
In ordinary gauge theories the gauge group and the set of fields can be chosen quite
independently and arbitrarily. The same is not true in the extended gauge models.  In order that
the gauge indices participate in the space-time transformations of the fields it is necessary that
the dimensionality of the fundamental matrix representation of the gauge group, say $d \times
d$, exactly matches the corresponding representations of the space-time symmetry group.  This
restricts greatly the possible choice of gauge group.  We make the simplest choice of gauge
group, namely $SO(4)$ itself.  The Hermitian generators of $SO(4)$ are, in the fundamental
representation,
$$\left( M^{\mu\nu}\right)^{\alpha\beta} = i\left(\delta^{\mu\alpha} \delta^{\nu\beta} -
\delta^{\mu\beta} \delta^{\nu\alpha}\right)\eqno(3)$$
satisfying
$$\left[M^{\mu\nu}, M^{\lambda\rho}\right] = i\left(\delta^{\nu\lambda} M^{\mu\rho}
+\delta^{\mu\rho} M^{\nu\lambda} - \delta^{\mu\lambda} M^{\nu\rho} - \delta^{\nu\rho}
M^{\mu\lambda} \right)\eqno(4)$$
and we will write the gauge field as
$$(A^\mu)^{\alpha\beta} = \sbody12 \,t^{\mu\nu\lambda}
(M^{\nu\lambda})^{\alpha\beta}\;\;\;\;\left(t^{\mu\nu\lambda} = -
t^{\mu\lambda\nu}\right).\eqno(5)$$
$$\hspace{-1cm} = i t^{\mu\alpha\beta}.\eqno(6)$$
Despite the occurrence of the factor of $i$ in (6), we note that the $t^{\mu\alpha\beta}$ are real
fields, as $(A^{\mu\dagger})^{\alpha\beta} = (A^{\mu})^{\beta\alpha *} = (i
t^{\mu\beta\alpha})^*
= i t^{\mu\alpha\beta}$. In terms of $t$ the field strength is, from (2), given by
$$ (F^{\mu\nu})^{\alpha\beta} = \sbody12 \left[ \partial^\mu t^{\nu\lambda\rho} - \partial^\nu
t^{\mu\lambda\rho} + \sbody12 \left(t^{\mu\tau\lambda} t^{\nu\tau\rho} - t^{\nu\tau\lambda}
t^{\mu\tau\rho}\right)\right] (M^{\lambda\rho})^{\alpha\beta}\eqno(7a)$$
$$  = i \left[ \partial^\mu t^{\nu\alpha\beta} - \partial^\nu t^{\mu\alpha\beta} + \sbody12
\left(t^{\mu\tau\alpha} t^{\nu\tau\beta} - t^{\nu\tau\alpha} t^{\mu\tau\beta}\right)\right].
\eqno(7b)$$

We now introduce Dirac 4-spinor matter fields in the fundamental representation of $SO(4)$
$\Psi^\alpha$.  For the matter fields we choose the following $SO(4)$ gauge invariant, and
Hermitian, Lagrangian density in $4dE$
$$L_{matter} = \Psi^\dagger (i\gamma \cdot \partial + \gamma \cdot A)\Psi\nonumber$$
$$\qquad\qquad = \Psi^{\alpha \dagger} (i\delta^{\alpha\beta} \gamma \cdot \partial + \gamma
\cdot
A^{\alpha\beta})\Psi^\beta.\eqno(8a)$$
$$\qquad\qquad = i\Psi^{\alpha \dagger} \gamma^\mu(\delta^{\alpha\beta}\partial^\mu +
t^{\mu\alpha\beta})\Psi^\beta.\eqno(8b)$$
We use Hermitian $\gamma$-matrices.  Our Euclidean space conventions are summarised in
Appendix A and are discussed in more detail in [3].
The action
$$S = \int d^4x \left(L_{YM} + L_{matter}\right)\eqno(9)$$
is clearly invariant under the usual local $SO(4)$ gauge transformations
$$\hspace{-1cm}\Psi(x) \longrightarrow U(\omega(x)) \Psi(x)\eqno(10a)$$
$$A^\mu(x) \longrightarrow U(\omega(x)) (A^\mu + i \partial^\mu) U^{-1}
(\omega(x))\eqno(10b)$$
where $U(\omega(x)) = \exp \left[ \frac{i}{2} \omega^{\lambda \rho} (x)
M^{\lambda\rho}\right]$. The
model of (9) is an $SO(4)$ gauge model.

The space-time symmetry group in $4dE$ is $SO(4)$. In our model we {\underline{extend}} the
usual field transformations under this $SO(4)$ group by allowing the gauge group indices
($\alpha\beta$ on $A^\mu$ and $\mu$ on $\Psi$) to participate in the transformations, as
follows,
$$\hspace{-2cm}(A^\mu)^{\alpha\beta} (x) \longrightarrow (A^{\prime\mu})^{\alpha\beta}
(x^\prime) = \Lambda^{\mu\nu} \left(U(\lambda)A^\nu (x) U^{-1} (\lambda)
\right)^{\alpha\beta}\nonumber$$
$$= \Lambda^{\mu\nu} U(\lambda)^{\alpha\rho} (A^\nu (x))^{\rho\sigma} U^{-
1}(\lambda)^{\sigma\beta}\eqno(11)$$
$$\hspace{-1.5cm}\Psi_i^\mu (x) \longrightarrow \Psi_i^{\prime\mu} (x^\prime) =
S_{ij}(\lambda)\left(U(\lambda)\Psi_j (x)\right)^\mu\nonumber$$
$$= S_{ij}(\lambda)U(\lambda)^{\mu\rho} \Psi_j^\rho(x)\eqno(12)$$
where
$$S_{ij}(\lambda) = \exp \left[\frac{i}{2} \lambda^{\mu\nu} \Sigma^{\mu\nu} \right]_{ij}\;
,\eqno(13)$$
$$U(\lambda)^{\alpha\rho} = \Lambda^{\alpha\rho} , \eqno(14)$$
$$x^{\prime\mu} = \Lambda^{\mu\nu} x^\nu\;\; .\eqno(15)$$
(Here, $\lambda^{\mu\nu} = -\lambda^{\nu\mu}$ is the matrix of four dimensional angles
parametrising the general four dimensional rotation matrix $\Lambda$; for the precise relationship
between $\lambda$ and $\Lambda$ see [4].)

It is clear that the gauge potential $t^{\mu\nu\lambda}$ introduced in (5) is reducible under
$SO(4)$, i.e. the gauge potential transforms, under the space-time $SO(4)$ transformations,
according to a non-irreducible representation.  (The use of non-irreducible representations of the
Lorentz group has previously been suggested by Dirac [5].)  Indeed, the matter field
$\Psi^\alpha$ is similarly $SO(4)$-reducible.  We now consider the decomposition of
$t^{\mu\nu\lambda}$ and $\Psi^\alpha$ into $SO(4)$-irreducible components.

The gauge potential $t^{\mu\nu\lambda}$ decomposes in terms of
\begin{itemize}
\item a vector field $v^\mu$
\item an axial vector field $a^\mu$
\item two rank 3 tensor fields $\Delta_{S,A}^{\mu\nu\lambda}$ which are
\begin{itemize}
\item[(1)] anti-symmetric in the final two indices:
$$\Delta_{S,A}^{\mu\nu\lambda} = - \Delta_{S,A}^{\mu\lambda\nu} \eqno(16)$$
\item[(2)] traceless over the first and either the second or third indices:
$$\Delta_{S,A}^{\mu\mu\lambda} = \Delta_{S,A}^{\mu\lambda\mu} = 0 \eqno(17)$$
\item[(3)] self-dual $S$ and anti-self dual $A$ respectively with respect to the final two indices:
$${^*}\Delta_{S,A}^{\mu\nu\lambda} = \sbody12\,\epsilon^{\nu\lambda\alpha\beta}
\Delta_{S,A}^{\mu\alpha\beta} = \pm \Delta_{S,A}^{\mu\nu\lambda}. \eqno(18)$$
\end{itemize}
\end{itemize}
The decomposition of the gauge potential is given by
$$t^{\mu\nu\lambda} = \delta^{\mu\nu} v^\lambda - \delta^{\mu\lambda} v^\nu +
\epsilon^{\mu\nu\lambda\rho} a^\rho + \Delta_S^{\mu\nu\lambda} +
\Delta_A^{\mu\nu\lambda}.\eqno(19)$$

The matter field 4-spinor $\Psi^\alpha$ decomposes in terms of 
\begin{itemize}
\item a Dirac $4$-spinor $\Psi$, and
\item an anti-symmetric tensor-spinor $\Psi^{\alpha\beta}$ where
\begin{itemize}
\item[(1)] $\Psi_R^{\alpha\beta} = \sbody12 (1 + \gamma_5)\Psi^{\alpha\beta}$ is self-dual
$$\hspace{-2cm}^* \Psi_R^{\alpha\beta} = \sbody12 \epsilon^{\alpha\beta\lambda\rho}
\Psi_R^{\lambda\rho}
= \Psi_R^{\lambda\rho}\eqno(20)$$
\item[(2)] $\Psi_L^{\alpha\beta} = \sbody12 (1 - \gamma_5)\psi^{\alpha\beta}$ is anti-self dual
$$\hspace{-2cm}^* \Psi_L^{\alpha\beta} = \sbody12 \epsilon^{\alpha\beta\lambda\rho}
\Psi_L^{\lambda\rho}
= - \Psi_L^{\alpha\beta}\eqno(21)$$
\end{itemize}
\end{itemize}
(together (20) and (21) imply that $^* \Psi^{\alpha\beta} = \gamma_5 \Psi^{\alpha\beta}$)
and the decomposition of the matter field is
$$\Psi^\alpha = \sbody12 \gamma^\alpha \Psi + \gamma^\lambda \Psi^{\lambda\alpha}
.\eqno(22)$$
The derivation of these results, (19) and (22), and the spin content of these $SO(4)$ irreducible
decompositions are treated in the following sections, and the spin
content of these $SO(4)$ irreducible decompositions discussed.

\section{Decomposition of Reducible Gauge Potential}

It is well known [6] that $SO(4) = SU(2) \times SU(2)$, and that the representations of
$SO(4)$ can be labelled by pairs of $SU(2)$ labels. The occurrence of half-odd-integer labels
for $SU(2)$ indicates spinor representations, eg the ($\sbody12, 0$) and ($0, \sbody12$)
representations correspond to the two fundamental inequivalent $2$-spinor representations of
$SO(4)$. The bispinor ($\sbody12 , \sbody12$) representation corresponds to the fundamental,
or 4-vector, representation of $SO(4)$.

The reducible gauge potential $t^{\mu\nu\lambda}$ involves a product of three ($\sbody12 , 
\sbody12$) representations.  In a product of two ($\sbody12 , \sbody12$) (representations)
$$(\sbody12 , \sbody12) \otimes (\sbody12 , \sbody12) = (1, 1) \oplus (1, 0) \oplus (0, 1) \oplus
(0, 0)\eqno(23)$$
the anti-symmetric combination of the two $4$-vector indices $\nu$ and $\lambda$ corresponds
to $(1, 0) \oplus (0, 1)$, while the symmetric combination corresponds to $(1, 1) \oplus (0, 0)$.
Consequently, corresponding to $t^{\mu\nu\lambda}$ is the product of representations
$$(\sbody12 , \sbody12) \otimes [(1, 0) \oplus (0, 1)] = (\sbody32 , \sbody12) \oplus (\sbody12
, \sbody12 ) \oplus (\sbody12 , \sbody12) \oplus (\sbody12 , \sbody32).\eqno(24)$$
Each of the representations on the right-hand side of (24) is irreducible.  This is the
decomposition that we are seeking.

The spin-content of these irreducible representations can be identified if we focus on the $SO(3)$
subgroup of $SO(4)$ corresponding to spatial rotations: $(\sbody32 , \sbody12)$ and $(\sbody12
, \sbody32)$ each contain spins one and two while $(\sbody12 , \sbody12)$ contains spin zero
and one.  Thus $t^{\mu\nu\lambda}$ will involve two spin-two states, four spin-one states and
two spin-zero states - in all twenty four independent degrees of freedom in agreement with the
expected number for $t^{\mu\nu\lambda} = - t^{\mu\lambda\nu}$.

To describe the decomposition of $t^{\mu\nu\lambda}$ corresponding to (24) it is necessary to
use the dotted and undotted spinor notation for $SO(4)$.  This notation is explained in detail in
Appendix
[3] and a summary of its salient features is provided in Appendix A.  It is crucial to note that
the
analysis differs considerably from the corresponding analysis for $SO(3, 1)$ in four dimensional
Minkowski space.

The first step in the decomposition is to write $t^{\mu\nu\lambda}$ as the sum of self-dual and
anti-self-dual parts
$$t^{\mu\nu\lambda} = t_S^{\mu\nu\lambda} + t_A^{\mu\nu\lambda}\eqno(25a)$$
where
$$t_S^{\mu\nu\lambda} = \sbody12 \left[ t^{\mu\nu\lambda} +
\,^*t^{\mu\nu\lambda}\right],\eqno(25b)$$
$$t_A^{\mu\nu\lambda} = \sbody12 \left[ t^{\mu\nu\lambda} -
\,^*t^{\mu\nu\lambda}\right],\eqno(25c)$$
with
$$^*t^{\mu\nu\lambda} = \sbody12 \,\epsilon^{\nu\lambda\alpha\beta} t^{\mu\alpha\beta}
.\eqno(25d)$$

The second step is to transform from the anti-symmetric $\nu,\lambda$ indices for the self-and
anti-self-dual parts of $t^{\mu\nu\lambda}$ to spinor indices (see Appendix B)
$$t_S^{\mu\nu\lambda} = -\sbody12 t_{S\;\;a}^{\;\mu\;\;\;b}
(\sigma^{\nu\lambda})_b^{\;\;a},\eqno(26a)$$
$$t_A^{\mu\nu\lambda} = -\sbody12 t_{A\;\;\;\;\dot{b}}^{\;\mu\;\;\dot{a}}
(\dot{\sigma}^{\nu\lambda})^{\dot{b}}_{\;\;\dot{a}}\eqno(26b)$$
where
$$t_{S\;a}^{\mu\;\;\;\;b} =
t_S^{\mu\nu\lambda}(\sigma^{\nu\lambda})_a^{\;\;\;b},\eqno(26c)$$
$$t_{A\;\;\;\;\dot{b}}^{\mu\;\;\dot{a}} =
t_A^{\mu\nu\lambda}(\dot{\sigma}^{\nu\lambda})^{\dot{a}}_{\;\;\;\dot{b}}.\eqno(26d)$$
Next, we transform from the remaining $4$-vector index $\mu$ to bi-spinor indices (see
Appendix B)
$$t_{S\;\;a}^{\mu\;\;\;\;b} = \sbody12\,(t_S)_{m\dot{n}a}^{\;\;\;\;\;\;
\;\;b}(\overline{\sigma}^\mu)^{\dot{n}m},\eqno(27a)$$
$$t_{A\;\;\;\;\dot{b}}^{\mu\;\;\dot{a}} =
\sbody12\,(t_A)^{\dot{m}n\dot{a}}_{\;\;\;\;\;\;\;\;\dot{b}}
(\sigma^\mu)_{n\dot{m}},\eqno(27b)$$
where
$$(t_S)_{m\dot{n}a}^{\;\;\;\;\;\;\;\;b} = t_{S\;\; a}^{\mu\;\;\;\;b}
(\sigma^\mu)_{m\dot{n}},\eqno(27c)$$
$$(t_A)^{\dot{m}n \dot{a}}_{\;\;\;\;\;\;\;\;\dot{b}} = t_{A\;\;\;\;\dot{b}}^{\mu\;\;\dot{a}}
(\overline{\sigma}^\mu)^{\dot{m}n}.\eqno(27d)$$
Putting these three steps together we find
$$t^{\mu\nu\lambda} = -\sbody14 \left[ (t_S)_{m\dot{n}a}^{\;\;\;\;\;\;\;b}
(\overline{\sigma}^\mu)^{\dot{n}m} (\sigma^{\nu\lambda})_b^{\;\;a} + (t_A)^{\dot{m} n
\dot{a}}_{\;\;\;\;\;\;\; \dot{b}}
(\sigma^\mu)_{n\dot{m}}(\dot\sigma^{\nu\lambda})^{\dot{b}}_{\;\;\dot{a}}\right]\nonumber$$
$$\quad = \sbody14 \left[ (t_S)_{m\dot{n}ab} (\overline{\sigma}^\mu)^{\dot{n}m}
(\sigma^{\nu\lambda})^{ba} + (t_A)^{\dot{m} n
\dot{a}\dot{b}}(\sigma^\mu)_{n\dot{m}}(\dot\sigma^{\nu\lambda})_{\dot{b}\dot{a}}\right]
.\eqno(28)$$
$(t_S)_{m\dot{n}ab}$ is explicitly symmetric in $(a,b)$ but has no particular symmetry with
respect to $(m,a)$ (or $(m,b)$).  Thus we can expand it in terms of the anti-symmetric matrix
$\epsilon_{ma}$ and the symmetric matrices $(\sigma^{\alpha\beta})_{ma}$
$$\hspace{-1cm}(t_S)_{m\dot{n}ab} = \sbody12 \left[p_{a\dot{n}} \epsilon_{mb} +
p_{b\dot{n}}
\epsilon_{ma} + (\Delta_S)_{a\dot{n}}^{\;\;\;\alpha\beta} (\sigma^{\alpha\beta})_{mb} +
(\Delta_S)_{b\dot{n}}^{\;\;\;\alpha\beta} (\sigma^{\alpha\beta})_{ma}\right]\nonumber$$
$$ = \sbody12 \left[p_{a\dot{n}} \epsilon_{mb} + p_{b\dot{n}} \epsilon_{ma} +
(\Delta_S)^{\mu\alpha\beta}\left( (\sigma^{\mu})_{a\dot{n}} (\sigma^{\alpha\beta})_{mb} +
(\sigma^\mu)_{b\dot{n}}(\sigma^{\alpha\beta})_{ma}\right)\right].\eqno(29)$$
In this expansion $p_{a\dot{n}}$ corresponds to the $(\sbody12 , \sbody12)$ part of
$(t_S)_{m\dot{n}ab}$ while $(\Delta_S)^{\mu\alpha\beta} \left( = \sbody12
(\Delta_S)_{a\dot{n}}^{\;\;\;\alpha\beta} (\overline{\sigma}^\mu)^{\dot{n}a}\right)$ corresponds
to the $(\sbody32 , \sbody12 )$ part.  However, the $(\sbody32 , \sbody12 )$ part must be totally
symmetric in the three spinor indices $(m, a, b)$; as the second term in (29) is not explicitly
symmetric in $(m, a)$ we impose this symmetry by contracting the term with $\epsilon^{am}$
and equating to zero
$$\Delta_S^{\mu\alpha\beta} \epsilon^{am} \left[ (\sigma^\mu)_{a\dot{n}}
(\sigma^{\alpha\beta})_{mb} + (\sigma^\mu)_{b\dot{n}} (\sigma^{\alpha\beta})_{ma} \right] =
0\eqno(30)$$
which immediately gives (see Appendix A)
$$0 = \Delta_S^{\mu\alpha\beta} (\sigma^{\alpha\beta}\sigma^\mu)_{b\dot{n}} = -\sbody12
\Delta_S^{\mu\alpha\beta} \left[ \delta^{\alpha\mu}\sigma^\beta - \delta^{\beta\mu} \sigma^\alpha
+ \epsilon^{\alpha\beta\mu\rho} \sigma^\rho \right]_{b\dot{n}}.\eqno(31)$$
But, as $\epsilon^{\mu\alpha\beta\rho} \Delta_S^{\mu\alpha\beta} = 2\;^*\Delta_S^{\mu\mu\rho}
= 2 \Delta_S^{\;\mu\mu\rho}$, we find
$$\Delta_S^{\;\mu\mu\rho} \sigma^\rho_{\;b\dot{n}} = 0\;\; .\eqno(32)$$
We conclude that the ($\sbody32 , \sbody12$) part of $(t_S)_{m\dot{n}ab}$ is identified by
imposing the tracelessness constraint
$$\Delta_S^{\;\mu\mu\rho} = 0\; .\eqno(33)$$
Using (29) in the first term of (28) allows us to identify the self-dual part of
$t^{\mu\nu\lambda}$ as
$$t_S^{\;\mu\nu\lambda} = \sbody14\left[ p^\rho(\sigma^\rho)_{a\dot{n}}
(\overline{\sigma}^\mu)^{\dot{n}b} (\sigma^{\nu\lambda})_b^{\;\; a} - 
\Delta_S^{\rho\alpha\beta} (\sigma^\rho)_{a\dot{n}} (\sigma^{\alpha\beta})_m^{\;\;\;b}
(\overline{\sigma}^\mu)^{\dot{n}m} (\sigma^{\nu\lambda})_b^{\;\; a}\right]\nonumber$$
$$ = \sbody14 \left( p^\rho \,Tr \left[\sigma^\rho \overline{\sigma}^\mu \sigma^{\nu\lambda}
\right] - \Delta_S^{\;\rho\alpha\beta} Tr \left[\sigma^\rho \overline{\sigma}^\mu
\sigma^{\alpha\beta} \sigma^{\nu\lambda}\right]\right).\eqno(34)$$
Evaluating the traces of the products of $\sigma$-matrices using the various identities provided
in Appendix A we find
$$t_S^{\;\mu\nu\lambda} = \sbody14 \left(\delta^{\mu\nu} p^\lambda - \delta^{\mu\lambda}
p^\nu + \epsilon^{\mu\nu\lambda\rho}p^\rho\right) + \sbody12 \left(\Delta_S^{\;\mu\nu\lambda}
-
 \Delta_S^{\;\nu\lambda\mu} - \Delta_S^{\;\lambda\mu\nu}\right).\eqno(35)$$
The term in (35) involving $\Delta_S$ can be simplified using the result (B.4) derived in
Appendix B, namely
$$\Delta_S^{\;\mu\nu\lambda} + \Delta_S^{\;\nu\lambda\mu} + \Delta_S^{\;\lambda\mu\nu} =
0\nonumber$$
to give
$$t_S^{\;\mu\nu\lambda} = \sbody14 \left(\delta^{\mu\nu} p^\lambda - \delta^{\mu\lambda}p^\nu
+ \epsilon^{\mu\nu\lambda\rho} p^\rho\right) + \Delta_S^{\;\mu\nu\lambda}\;\;.\eqno(37)$$
We note that $\Delta_S^{\;\mu\nu\lambda}$ has $\sbody12 (4.6) - 4 = 8$ degrees of freedom -
the self-duality condition giving rise to the factor $\sbody12$ and the tracelessness constraint
giving rise to the $- 4$ - appropriate to the $(\sbody32 , \sbody12 )$ representation.

The anti-self-dual part of $t^{\mu\nu\lambda}$ can be expanded in an analogous manner.  In that
case we find
$$t_A^{\;\mu\nu\lambda} = \sbody14 \left(\delta^{\mu\nu} q^\lambda - \delta^{\mu\lambda}
q^\nu - \epsilon^{\mu\nu\lambda\rho} q^\rho \right) + \Delta_A^{\;\mu\nu\lambda}\eqno(38)$$
where $q^\lambda$ belongs to the $(\sbody12 , \sbody12)$ representation and
$\Delta_A^{\;\mu\nu\lambda}$ is anti-self-dual
$$^*\Delta_A^{\mu\nu\lambda} = \sbody12 \epsilon^{\nu\lambda\alpha\beta}
\Delta_A^{\;\mu\alpha\beta} = -\Delta_A^{\;\mu\nu\lambda} \; ,\eqno(39a)$$
satisfies
$$\Delta_A^{\mu\mu\rho} = 0,\eqno(39b)$$
and belongs to the $(\sbody12 , \sbody32 )$ representation.

Combining the results (37) and (38) for $t_S$ and $t_A$ we finally obtain for the decomposition
of the reducible gauge potential in terms of irreducible $SO(4)$ components
$$t^{\mu\nu\lambda} = \delta^{\mu\nu} v^\lambda - \delta^{\mu\lambda} v^\nu +
\epsilon^{\mu\nu\lambda\rho} a^\rho + \Delta_S^{\;\mu\nu\lambda} +
\Delta_A^{\;\mu\nu\lambda}\eqno(40)$$
where $v^\lambda = \sbody14 (p + q)^\lambda$ and $a^\lambda = \sbody14 (p - q)^\lambda$.
We note that the decomposition derived here is somewhat different from that proposed in [1].

To assist in the analysis of the gauge transformations of the fields in Section 5 we indicate now
the projection from the reducible gauge potential to each of its irreducible components.  The
projection for the vector field is straightforward
$$v^\lambda = \sbody13 t^{\mu\mu\lambda}\; .\eqno(41)$$
The projection for the axial vector field can be similarly written, noting the dual of (40) above,
namely
$$^*t^{\mu\nu\lambda} = \delta^{\mu\nu} a^\lambda - \delta^{\mu\lambda} a^\nu +
\epsilon^{\mu\nu\lambda\rho} v^\rho + \Delta_S^{\;\mu\nu\lambda} -
\Delta_A^{\;\mu\nu\lambda}\eqno(42)$$
so that the roles of $v$ and $a$ are interchanged in $^*t^{\mu\nu\lambda}$.  The projection is
$$a^\lambda = \sbody13\, ^*t^{\mu\mu\lambda} = \sbody16\, \epsilon^{\mu\beta\gamma\lambda}
t^{\mu\beta\gamma}\; .\eqno(43)$$
We introduce two ``orthogonal'' linear combinations of the fields $\Delta_S$ and $\Delta_A$,
namely
$$\Delta_{\pm}^{\;\mu\nu\lambda} = \Delta_S^{\;\mu\nu\lambda} \pm
\Delta_A^{\;\mu\nu\lambda}\eqno(44)$$
which are dual to one another
$$^*\Delta_+^{\mu\nu\lambda} = \sbody12\, \epsilon^{\nu\lambda\alpha\beta}
\Delta_+^{\mu\alpha\beta} = \Delta_-^{\mu\nu\lambda}\;\; .\eqno(45)$$
The projection from $t$ to $\Delta_+$ is found by using (41) and (43) above in
$$\Delta_+^{\;\mu\nu\lambda} = t^{\mu\nu\lambda} - \left( \delta^{\mu\nu} v^\lambda -
\delta^{\mu\lambda} v^\nu + \epsilon^{\mu\nu\lambda\rho} a^\rho\right)\nonumber$$
to find
$$\Delta_+^{\;\mu\nu\lambda} = \sbody13 \left[ 2t^{\mu\nu\lambda} - t^{\lambda\mu\nu} -
t^{\nu\lambda\mu} - \delta^{\mu\nu} t^{\alpha\alpha\lambda} +
\delta^{\mu\lambda}t^{\alpha\alpha\nu}\right]\; .\eqno(46)$$
The relationship between $^*t$ and $\Delta_-$ evident in (42) above leads to a similar projection
from $^*t$ to $\Delta_-$, namely
$$\Delta_-^{\;\mu\nu\lambda} = \sbody13 \left[ 2\,^*t^{\mu\nu\lambda} -\, ^*t^{\lambda\mu\nu}
-
 ^*t^{\nu\lambda\mu} - \delta^{\mu\nu}\; ^*t^{\alpha\alpha\lambda} + \delta^{\mu\lambda}\;
^*t^{\alpha\alpha\nu} \right].\eqno(47)$$
Combining (46) and (47) we now find the projections to $\Delta_{S,A}$
$$\Delta_{S,A}^{\;\;\mu\nu\lambda} = \sbody16 \left[2(t^{\mu\nu\lambda} \pm
\,^*t^{\mu\nu\lambda}) - (t^{\lambda\mu\nu} \pm \, ^* t^{\lambda\mu\nu}) - (t^{\nu\lambda\mu}
\pm \, ^*t^{\nu\lambda\mu})\right.\nonumber$$
$$\left. - \delta^{\mu\nu} (t^{\alpha\alpha\lambda} \pm  \, ^*t^{\alpha\alpha\lambda}) +
\delta^{\mu\lambda}(t^{\alpha\alpha\nu} \pm \, ^*t^{\alpha\alpha\nu})\right]\eqno(48)$$
where the $+$ sign is used for $\Delta_S$ and the $-$ sign for $\Delta_A$.

\section{Decomposition of Reducible Matter Spinor Fields}

The matter $4$-spinor field introduced in Section 2 $\Psi^\mu_i$ carries a $4$-vector index in
that
participates in both the $SO(4)$ gauge and $SO(4)$ Euclidean space transformation, and a Dirac
spinor index that participates only in the $SO(4)$ Euclidean space transformation.  Thus it is
associated with the reducible product of representations
$$(\sbody12 , \sbody12 ) \otimes [(\sbody12 , 0) \oplus (0, \sbody12 )] = (1, \sbody12 ) \oplus
(0, \sbody12 ) \oplus (\sbody12 , 0) \oplus (\sbody12 , 1).\eqno(49)$$
Each of the representations on the right hand side of (49) is $SO(4)$-irreducible.  To understand
this decomposition it is necessary to use the $2$-spinor notation in terms of which
$$\Psi_i^\mu = \left[ \begin{array}{c} \psi_a^\mu \\ \chi^{\dot{a}\mu}\end{array} \right]\;\;
.\eqno(50)$$
We will focus separately on the undotted and dotted spinor components. The spinors
$\psi_a^\mu$, $\chi^{\dot{a}\mu}$ correspond respectively to the reducible representations
$$(\sbody12 , \sbody12 ) \otimes (\sbody12 , 0) = (0, \sbody12 ) \oplus (1, \sbody12
),\eqno(51a)$$
$$(\sbody12 , \sbody12 ) \otimes (0, \sbody12 ) = (\sbody12 , 0 ) \oplus (\sbody12 ,
1).\eqno(51b)$$
As the $(1, \sbody12 )$ and $(\sbody12 , 1 )$ representations each contain a spin $\sbody32$ and
spin $\sbody12$ state we see that each of $\psi_a^\mu$ and $\chi^{\dot{a}\mu}$ contain one
spin $\sbody32$ state and two spin $\sbody12$ states.  This corresponds to $2.(4 + 2.2) = 16$
states, appropriate to the vector-spinor field $\Psi^\mu$.

We first transform the vector index on $\psi_a^\mu$ to bispinor indices as follows
$$\psi_a^\mu = \sbody12 \psi_a^{\;\;\;\dot{b}c} (\sigma^\mu)_{a\dot{b}} = \sbody12
\psi_{a\;\;\;\;d}^{\;\;\;\dot{b}} \epsilon^{cd}
(\sigma^\mu)_{c\dot{b}}\eqno(52a)$$
where
$$\psi_a^{\;\;\;\dot{b}c} = \psi_a^\mu(\overline{\sigma}^\mu)^{\dot{b}c}.\eqno(52b)$$
The $(0, \sbody12 )$ component corresponds to that part of $\psi_{a\;\;\;d}^{\;\;\dot{b}}$ 
anti-symmetric in $(a,b)$ while the $(1, \sbody12 )$ corresponds to the symmetric part.  Thus
we have the expansion
$$\psi_{a\;\;\;d}^{\;\;\;\dot{b}} = -\psi^{\dot{b}} \epsilon_{ad} + (\psi_S^{\alpha\beta}
)^{\dot{b}}
(\sigma^{\alpha\beta})_{ad}\eqno(53a)$$
where
$$^*\psi_S^{\alpha\beta} = \sbody12\, \epsilon^{\alpha\beta\rho\sigma} \psi_S^{\rho\sigma} =
\psi_S^{\alpha\beta}\eqno(53b)$$
is self-dual.  This leads at once to the decomposition of $\psi_a^\mu$ in terms of its 
$SO(4)$-irreducible components
$$\psi_a^\mu = \sbody12\, \sigma_{\;\;a\dot{b}}^\mu \psi^{\dot{b}} - \sigma_{\;a\dot{b}}^\alpha
(\psi_S^{\mu\alpha})^{\dot{b}}\; .\eqno(54)$$

The projection from $\psi_a^\mu$ to the $(0, \sbody12 )$ spinor $\psi^{\dot{b}}$ is 
$$\psi^{\dot{b}} = \sbody12 (\overline{\sigma}^\mu)^{\dot{b}c} \psi_c^{\;\mu}\; .\eqno(55)$$
To derive the appropriate projection from $\psi_a^\mu$ to the $(1, \sbody12 )$ spinor
$(\psi_S^{\alpha\beta})^{\dot{b}}$ we use (55) in (53a) to find
$$(\psi_S^{\alpha\beta})^{\dot{b}} (\sigma^{\alpha\beta})_{ad} = - \psi_{a\;\;d}^{\,\;\dot{b}}
+
\sbody12 (\overline{\sigma}^\mu)^{\dot{b}c} \psi_c^\mu \epsilon_{ad}\; .\eqno(56)$$
Raising the $d$ index in this equation and using (52b) to relate
$\psi_a^{\;\;\;\dot{b}c}$ to $\psi_a^\mu$ we obtain
$$(\psi_S^{\alpha\beta})^{\dot{b}}(\sigma^{\alpha\beta})_a^{\;\;\;c} = -
\psi_a^\mu(\overline{\sigma}^\mu)^{\dot{b}c} - \sbody12(\overline{\sigma}^\mu)^{\dot{b}e}
\psi_e^\mu \delta_a^{\;\;\;c}.\eqno(57)$$
Multiplying this equation by $(\sigma^{\mu\nu})_c^{\;\;\;d}$ and tracing over the $(a,d)$ indices
leads to the projection
$$\hspace{-1cm}(\psi_S^{\mu\nu})^{\dot{b}} = \sbody14\left[\delta^{\mu\alpha}\delta^{\nu\beta}
- \delta^{\mu\beta}\delta^{\nu\alpha} + \epsilon^{\mu\nu\alpha\beta} \right]
(\overline{\sigma}^\alpha)^{\dot{b}a} \psi_a^\beta\eqno(58a)$$
$$= \sbody14\left[(\overline{\sigma}^\mu)^{\dot{b}a} \psi_a^\nu
- (\overline{\sigma}^\nu)^{\dot{b}a}\psi_a^\mu
+ \epsilon^{\mu\nu\alpha\beta}(\overline{\sigma}^\alpha)^{\dot{b}a} \psi_a^\beta
\right].  \eqno(58b)$$

The decomposition of $\chi^{\dot{a}\mu}$ into $( \sbody12 , 0)$ and $( \sbody12 , 1)$ states
proceeds in a very similar manner.  We first transform to the spinor indices
$$\chi^{\dot{a}\mu} = \sbody12 \chi^{\dot{a}}_{\;\;b\dot{c}}
(\overline{\sigma}^\mu)^{\dot{c}b} = \sbody12 \chi^{\dot{a}\;\;\dot{d}}_{\,\;b}
\epsilon_{\dot{c}\dot{d}}(\overline{\sigma}^\mu)^{\dot{c}b}\;\; ,\eqno(59)$$
and then expand $\chi^{\dot{a}\;\;\dot{d}}_{\,\;b}$ in terms of a part anti-symmetric in $(\dot{a}
, \dot{d})$ and an anti-self-dual part symmetric in $(\dot{a} , \dot{d})$
$$\chi_{\,\;b}^{\dot{a}\;\;\dot{d}} = -\chi_b \epsilon^{\dot{a} \dot{d}} +
\left(\chi_A^{\alpha\beta}\right)_b (\dot{\sigma}^{\alpha\beta})^{\dot{a}\dot{d}}\eqno(60)$$
giving us for the decomposition
$$\chi^{\dot{a}\mu} = \sbody12 (\overline{\sigma}^\mu)^{\dot{a}b}
\chi_b - (\overline{\sigma}^\alpha)^{\dot{a}b}
(\chi_A^{\mu\alpha})_b.\eqno(61)$$
The projection of $\chi^{\dot{a}\mu}$ to the two $SO(4)$ irreducible components are, then, 
$$\hspace{-2cm}\chi_a = \sbody12 (\sigma^\mu)_{a\dot{b}} \chi^{\dot{b}\mu}\eqno(62a)$$
$$\left(\chi_A^{\mu\nu}\right)_a = \sbody14 \left[(\sigma^\mu)_{a\dot{b}}\chi^{\dot{b}\nu}
 - (\sigma^\nu)_{a\dot{b}}\chi^{\dot{b}\mu} - 
\epsilon^{\mu\nu\alpha\beta} (\sigma^\alpha)_{a\dot{b}} \chi^{\dot{b}\beta}
\right].\eqno(62b)$$
The decompositions (54) and (61) can be combined together to give the $SO(4)$ decomposition
of the vector-spinor matter field $\Psi^\mu$ in terms of a Dirac $4$-spinor
$$\Psi = \left[ \begin{array}{c} \chi_a \\ \psi^{\dot{a}} \end{array}\right]\eqno(63a)$$
and a tensor-spinor
$$\Psi^{\alpha\beta} = \left[ \begin{array}{c} \chi_{A\;\;a}^{\;\alpha\beta} \\ 
\psi_{S}^{\alpha\beta\dot{a}} \end{array}\right].\eqno(63b)$$
We find
$$\Psi^\mu = \sbody12 \gamma^\mu \Psi + \gamma^\alpha \Psi^{\alpha\mu} \; .\eqno(63c)$$
It is clear that the right chirality part of $\Psi^{\alpha\beta}$is self-dual
$$\Psi_R^{\alpha\beta} = \sbody12 (1 + \gamma_5) \Psi^{\alpha\beta} = \left[ \begin{array}{c}
0 \\ \psi_S^{\alpha\beta\dot{a}}\end{array}\right] \eqno(64a)$$
while the left chirality part of $\Psi^{\alpha\beta}$ is anti-self-dual
$$\Psi_L^{\alpha\beta} = \sbody12 ( 1- \gamma_5) \Psi^{\alpha\beta} =
\left[ \begin{array}{c} \chi_{A\;\;a}^{\alpha\beta}\\ 0\end{array}\right], \eqno(64b)$$
so that $^*\Psi^{\alpha\beta} = \gamma_5 \Psi^{\alpha\beta}$.
The projection from $\Psi^\mu$ to $\Psi$ and $\Psi^{\alpha\beta}$, or to their right- and left-
chirality parts, can be written in $4$-component form as
$$\Psi = \sbody12 \gamma^\mu \Psi^\mu\eqno(65a)$$
$$\Psi_R = \frac{1 + \gamma_5}{2} \Psi = \sbody12 \gamma^\mu  \Psi_L^\mu\eqno(65b)$$
$$\Psi_L = \frac{1 - \gamma_5}{2} \Psi = \sbody12 \gamma^\mu \Psi_R^\mu\eqno(65c)$$ 
and
$$\Psi^{\alpha\beta} = \sbody14 \left( \gamma^\alpha \Psi^\beta - \gamma^\beta \Psi^\alpha +
\epsilon^{\alpha\beta\lambda\rho} \gamma_5 \gamma^\lambda \Psi^\rho \right)\eqno(66a)$$
$$\Psi_R^{\alpha\beta} = \sbody14 \left( \gamma^\alpha \Psi_L^\beta - \gamma^\beta
\Psi_L^\alpha + \epsilon^{\alpha\beta\lambda\rho} \gamma_5 \gamma^\lambda \Psi_L^\rho
\right)\eqno(66b)$$
$$\Psi_L^{\alpha\beta} = \sbody14 \left( \gamma^\alpha \Psi_R^\beta - \gamma^\beta
\Psi_R^\alpha + \epsilon^{\alpha\beta\lambda\rho} \gamma_5 \gamma^\lambda \Psi_R^\rho
\right).\eqno(66c)$$

In $4dE$ the adjoint of the matter vector-spinor field is the Hermitian conjugate
$\Psi^{\mu\dagger}$. The conjugate of the decomposition (63c) is simply
$$\Psi^{\mu\dagger} = \sbody12 \Psi^\dagger \gamma^\mu - \Psi^{\mu\alpha{^{\dagger}}}
\gamma^\alpha \eqno(67)$$
as the Dirac $\gamma$-matrices have been chosen to be Hermitian.  In terms of the 
$SU(2)\;2$-spinors these fields are
$$\Psi^{\mu\dagger} = \left[\overline{\psi}^{\mu\alpha}\; ,  - \overline{\chi}^\mu_{\;\dot{a}}
\right]\; , \eqno(68a)$$
$$\Psi^\dagger = \left[\overline{\chi}^b\; ,  - \overline{\psi}_{\dot{b}} \right]\; , \eqno(68b)$$
$$\Psi^{\mu\alpha{^{\dagger}}} = \left[\overline{\chi}_A^{\mu\alpha b}\; , -
\overline{\psi}_{S\;\;\dot{b}}^{\mu\alpha}\right]\; .\eqno(68c)$$
The decompositions in terms of $SO(4)$-irreducible components are found by taking
Hermitian
conjugates of (54) and (61)
$$\overline{\psi}^{\mu a} = -\sbody12 \overline{\psi}_{\dot{b}}
(\overline{\sigma}^\mu)^{\dot{b} a} + \overline{\psi}_{S\;\;\dot{b}}^{\mu\alpha}
(\overline{\sigma}^\alpha)^{\dot{b}a},\eqno(69a)$$
$$\overline{\chi}^\mu_{\dot{a}} = -\sbody12 \overline{\chi}^b (\sigma^\mu)_{b\dot{a}} +
\overline{\chi}_A^{\mu\alpha\; b} (\sigma^\alpha)_{b\dot{a}}.\eqno(69b)$$
In taking Hermitian conjugates of $2$-spinors in $4dE$ we must pay particular attention to the
index structure of the fields involved.  The rules are derived in [3] and summarised in Appendix
A.

\section{Gauge Transformations and Gauge Invariant Lagrangian in terms of
Irreducible Fields.}
\indent The $SO(4)$ gauge model with extended field transformations was described in Section
2 in
terms of the reducible gauge potential $A^\mu$ and vector-spinor matter field
$\Psi^\mu$.  Using
the results of Sections 3 and 4 we now consider the $SO(4)$ gauge transformations (10a,b) and
the $SO(4)$ gauge invariant action (9) in terms of the irreducible fields.

Equations (10a,b) are the finite gauge transformations.  In this section we restrict our attention
to the transformations for which the gauge parameters $\omega^{\alpha\beta} (x) (= -
\omega^{\beta\alpha} (x))$ are infinitesimal.  The corresponding gauge transformations for the
reducible gauge potential $t^{\mu\alpha\beta}$ can be written in three equivalent ways
$$\;\;\;\;\;\delta t^{\mu\alpha\beta} = \left\lbrace \begin{array}{l}
\partial^\mu \omega^{\alpha\beta} + t^{\mu\alpha\rho}\omega^{\rho\beta} -
t^{\mu\beta\rho}\omega^{\rho\alpha} \hspace{8.2cm}{\rm{(70{\it{a}})}}\\
\partial^\mu \omega^{\alpha\beta} + \epsilon^{\alpha\beta\lambda\rho}\,
{^*}t^{\mu\lambda\nu}\omega^{\nu\rho} \hspace{9cm}{\rm{(70{\it{b}})}}\\
\partial^\mu \omega^{\alpha\beta} + {^*}t^{\mu\alpha\rho}\,{^*}\omega^{\rho\beta} -
\,{^*}t^{\mu\beta\rho}{^*}\omega^{\rho\alpha}.
\hspace{7.4cm}{\rm{(70{\it{c}})}}
\end{array} \right.$$
The equivalence of these three transformations, and the equivalence in a number of other cases
below, can easily be established using the result (B.5) in Appendix B.  A similar gauge
transformation can be established for ${^*}t^{\mu\alpha\beta}$

$$\;\;\;\;\;\;\;\;\;\;\;\;\;\;\;\;\;\;\;\;\;\;\;\;\delta \,{^*}t^{\mu\alpha\beta} = \left\lbrace \begin{array}{l}
\partial^\mu \,{^*}\omega^{\alpha\beta} + t^{\mu\alpha\rho} \,{^*}\omega^{\rho\beta} -
t^{\mu\beta\rho}\,{^*}\omega^{\rho\alpha}\hspace{5.3cm}
{\rm{(71{\it{a}})}}\\
\partial^\mu \,{^*}\omega^{\alpha\beta} + \epsilon^{\alpha\beta\lambda\rho}\,
t^{\mu\lambda\nu}\,\omega^{\nu\rho} \hspace{6.6cm}{\rm{(71{\it{b}})}}\\
\partial^\mu \,{^*}\omega^{\alpha\beta} + {^*}t^{\mu\alpha\rho}\,\omega^{\rho\beta} -
{^*}t^{\mu\beta\rho}\,\omega^{\rho\alpha}.\hspace{5.2cm}
{\rm{(71{\it{c}})}}
\end{array} \right.$$
In a very obvious way we can deduce the following equivalent gauge transformations for the
vector field $v^\mu$ and the axial vector field $a^\mu$
$$\;\;\;\;\;\;\;\;\;\;\;\;\;\;\;\;\;\delta v^\mu = \left\lbrace \begin{array}{l} \sbody13 \left[\partial^\alpha
\omega^{\alpha\mu}
+ 2v^\alpha \omega^{\alpha\mu} + 2a^\alpha \,{^*}\omega^{\alpha\mu} +
\Delta_+^{\alpha\mu\beta}
\omega^{\alpha\beta} \right]\hspace{4.4cm} {\rm{(72{\it{a}})}}\\
\sbody13 \left[\partial^\alpha \omega^{\alpha\mu} + 2v^\alpha \omega^{\alpha\mu} + 2a^\alpha
\,{^*}\omega^{\alpha\mu} + \Delta_-^{\alpha\mu\beta} \,{^*}\omega^{\alpha\beta}
\right]\hspace{4.2cm}
{\rm{(72{\it{b}})}}\end{array}\right.$$
$$\;\;\;\;\;\;\;\;\;\;\;\;\;\;\;\;\;\;\delta a^\mu = \left\lbrace \begin{array}{l} \sbody13
\left[\partial^{\alpha}\,{^*}
\omega^{\alpha\mu} + 2v^{\alpha}\,{^*} \omega^{\alpha\mu} + 2a^\alpha \omega^{\alpha\mu}
+
\Delta_+^{\alpha\mu\beta} \,{^*}\omega^{\alpha\beta}
\right]\hspace{3.8cm} {\rm{(73{\it{a}})}}\\
\sbody13 \left[\partial^{\alpha}\,{^*} \omega^{\alpha\mu} + 2v^{\alpha}\,{^*}
\omega^{\alpha\mu} +
2a^\alpha \omega^{\alpha\mu} + \Delta_-^{\alpha\mu\beta}
\omega^{\alpha\beta} \right].\hspace{4cm}
{\rm{(73{\it{b}})}}\end{array}\right.$$

The gauge transformations for $\Delta_{\pm}^{\mu\nu\lambda}$ are more tedious to derive; we
give here just one of the equivalent forms in each case.  The transformations are
$$\delta \Delta_+^{\mu\nu\lambda} = \sbody13 \left\lbrace \left[ (\partial^\mu -
v^\mu)\omega^{\nu\lambda} - a^\mu \,{^*}\omega^{\nu\lambda}\right] - \left[ (\partial^\nu -
v^\nu)\omega^{\lambda\mu} - a^\nu\,{^*}\omega^{\lambda\mu}\right. \right]\nonumber$$
$$- \delta^{\mu\nu} \left[ (\partial^\alpha - v^\alpha)\omega^{\alpha\lambda} - a^\alpha
\,{^*}\omega^{\alpha\lambda}\right] - \delta^{\mu\nu} \Delta_+^{\alpha\lambda\beta}
\omega^{\alpha\beta}\nonumber$$
$$\left. + \left( \Delta_+^{\mu\nu\alpha} + \Delta_+^{\nu\mu\alpha}
\right)\omega^{\alpha\lambda} + \Delta_+^{\lambda\nu\alpha} \omega^{\alpha\mu}
\right\rbrace - (\nu \leftrightarrow \lambda)\eqno(74)$$

$$\delta \Delta_-^{\mu\nu\lambda} = \sbody13 \left\lbrace \left[ (\partial^\mu -
v^\mu)\,{^*}\omega^{\nu\lambda} - a^\mu \omega^{\nu\lambda}\right] - \left[ (\partial^\nu -
v^\nu)\,{^*}\omega^{\lambda\mu} - a^\nu \omega^{\lambda\mu}\right. \right]\nonumber$$
$$- \delta^{\mu\nu} \left[ (\partial^\alpha - v^\alpha)\,{^*}\omega^{\alpha\lambda} - a^\alpha
\omega^{\alpha\lambda}\right] - \delta^{\mu\nu} \Delta_+^{\alpha\lambda\beta}
\,{^*}\omega^{\alpha\beta}\nonumber$$
$$\left. + \left( \Delta_+^{\mu\nu\alpha} + \Delta_+^{\nu\mu\alpha}
\right){^*}\omega^{\alpha\lambda} + \Delta_+^{\lambda\nu\alpha} \, ^*\omega^{\alpha\lambda}
\right\rbrace - (\nu \leftrightarrow \lambda).\eqno(75)$$
These transformations can be used to find the gauge transformations of the $SO(4)$ irreducible
fields $\Delta_{S,A}$
$$\!\!\!\!\!\!\!\delta \Delta_{S,A}^{\mu\nu\lambda} = \sbody16 \left\lbrace \left(\partial^\mu -
v^\mu -
a^\mu\right) \left(\omega^{\nu\lambda} \pm {^*}\omega^{\nu\lambda}\right) - \left( \partial^\nu
- v^\nu - a^\nu\right) \left(\omega^{\lambda\mu} \pm
{^*}\omega^{\lambda\mu}\right)\right.
\nonumber$$
$$\hspace{2cm}- \delta^{\mu\nu} \left(\partial^\alpha - v^\alpha - a^\alpha\right)
\left(\omega^{\alpha\lambda}
\pm {^*}\omega^{\alpha\lambda}\right) - \delta^{\mu\nu} (\Delta_S +
\Delta_A)^{\alpha\lambda\beta} \left(\omega^{\alpha\beta} \pm
{^*}\omega^{\alpha\beta}\right)\nonumber$$
$$+ \left[( \Delta_S + \Delta_A)^{\mu\nu\alpha} + (\Delta_S + \Delta_A)^{\nu\mu\alpha} \right]
\left(\omega^{\alpha\lambda} \pm {^*}\omega^{\alpha\lambda} \right) \nonumber$$
$$+ \left.\left( \Delta_S + \Delta_A\right)^{\lambda\nu\alpha} \left(\omega^{\alpha\mu} \pm
{^*}\omega^{\alpha\mu} \right) \right\rbrace - (\nu \leftrightarrow \lambda)\eqno(76)$$
where the $+$ sign is used for $\delta\Delta_S$ and the $-$ sign for $\delta\Delta_A$.  By
inspection of the $SO(4)$ gauge transformations (72) - (76) we see that the gauge
transformations {\underline{mix}} together the different $SO(4)$ irreducible fields.  For instance
in eq. (72a) the vector $v^\mu$ is mixed with the axial vector $a^\mu$ and the $\Delta_+$ field,
which contains both spin $2$ and spin $1$ components.

In a similar vein, using the projections derived in the previous section from the reducible matter
field $\Psi^\mu$ to the ``irreducible'' $4$-spinors $\Psi$ and $\Psi^{\alpha\beta}$ we can deduce
from eq. (10a) the following $SO(4)$ gauge transformations
$$\delta \Psi = \sbody{i}{2} \omega^{\mu\nu} \Sigma^{\mu\nu} \Psi - \left[ i
\Sigma^{\mu\alpha}
\omega^{\alpha\nu} + \omega^{\mu\nu} \right]\Psi^{\mu\nu}\eqno(77)$$

$$\delta \Psi^{\alpha\beta} = \left( \omega^{\alpha\beta} + {^*}\omega^{\alpha\beta}\gamma_5
\right)\Psi - \sbody12 \epsilon^{\alpha\beta\lambda\nu} \left({^*}\omega^{\lambda\mu}  +
\gamma_5 \omega^{\lambda\mu})\Psi^{\mu\nu}\right)\nonumber$$
$$+ \sbody{i}{2} \left[ \left( \Sigma^{\alpha\mu} \omega^{\beta\nu} -
\Sigma^{\beta\mu} \omega^{\alpha\nu} \right) + 
\epsilon^{\alpha\beta\lambda\rho} \gamma_5 \Sigma^{\lambda\mu} \omega^{\rho\nu}
\right]\Psi^{\mu\nu}.\eqno(78)$$
Again, we see the inevitable mixing of $\Psi$ and $\Psi^{\mu\nu}$ in the gauge transformations.

The Lagrangian density is very simple in terms of the reducible fields as seen in $L_{YM}$ and
$L_{matter}$-eqs. (1) and (8).  In terms of the irreducible fields, however, we find a large
number of induced couplings.  It must be emphasized that the resulting Lagrangian density,
though complicated, is gauge invariant under the gauge transformations derived above.  We now
examine $L_{YM}$ and $L_{matter}$ separately.

Using the decomposition of the reducible gauge potential $t^{\mu\nu\lambda}$ derived in Section
3 in the field strength $(F^{\mu\nu})^{\alpha\beta}$ of eq. (7b), we can see that the pure 
Yang-Mills Lagrangian density is
$$L_{YM} = \sbody14 \left[ X_{(1)}^{\mu\nu\alpha\beta} + X_{(2)}^{\mu\nu\alpha\beta} - (\mu
\leftrightarrow \nu) \right]^2\eqno(79)$$
where $X_{(1)}^{\mu\nu\alpha\beta}$ is linear in the irreducible fields
$$X_{(1)}^{\mu\nu\alpha\beta} = \delta^{\nu\alpha} \partial^\mu v^\beta - \delta^{\nu\beta}
\partial^\mu v^\alpha + \epsilon^{\nu\alpha\beta\lambda} \partial^\mu a^\lambda + \partial^\mu
\Delta_+^{\nu\alpha\beta}\eqno(80)$$
and $X_{(2)}^{\mu\nu\alpha\beta}$ is quadratic in the irreducible fields
$$\!\!\!\!\!\!\!\!\!\!\!\!X_{(2)}^{\mu\nu\alpha\beta} = \sbody12\left[ \delta^{\mu\beta} v^\nu
v^\alpha -
\delta^{\mu\alpha} v^\nu v^\beta + \delta^{\mu\alpha} \delta^{\nu\beta} (v \cdot
v)\right.\nonumber$$
$$\hspace{3cm} + \delta^{\mu\beta} a^\nu a^\alpha - \delta^{\mu\alpha} a^\nu a^\beta +
\delta^{\mu\alpha}
\delta^{\nu\beta} (a \cdot a) + \Delta_+^{\mu\lambda\alpha}
\Delta_+^{\nu\lambda\beta}\nonumber$$
$$ \hspace{4cm}+ \left( \epsilon^{\mu\nu\alpha\rho} v^\beta - \epsilon^{\mu\nu\beta\rho}
v^\alpha\right) a^\rho
+ \left( \delta^{\mu\beta} \epsilon^{\alpha\nu\lambda\rho} - \delta^{\mu\alpha}
\epsilon^{\beta\nu\lambda\rho}\right) v^\lambda a^\rho \nonumber$$
$$\hspace{3cm}+ \left( \Delta_+^{\mu\nu\alpha} v^\beta -  \Delta_+^{\mu\nu\beta} v^\alpha
\right) + 
\left( \delta^{\mu\beta} \Delta_+^{\nu\lambda\alpha} - \delta^{\mu\alpha}
\Delta_+^{\nu\lambda\beta}\right) v^\lambda\nonumber$$
$$\left.\hspace{1cm} + \left( \Delta_+^{\mu\lambda\alpha} \epsilon^{\nu\lambda\beta\rho}
 -\Delta_+^{\mu\lambda\beta} \epsilon^{\nu\lambda\alpha\rho}\right)a^\rho \right]. \eqno(81)$$
We refrain from expanding out $L_{YM}$ in full detail satisfying ourselves instead with the
kinetic part which is quadratic in the fields
$$\hspace{-4cm}L_{YM}^{(2)} = \sbody14 \left( X_{(1)}^{\mu\nu\alpha\beta}
\right)^2\nonumber$$
$$ \hspace{4cm}= -v^\mu (2 \Box \delta^{\mu\nu} + \partial^\mu \partial^\nu)v^\nu - a^\mu (2
\Box
\partial^{\mu\nu} + \partial^\mu \partial^\nu)a^\nu\nonumber$$
$$\hspace{3cm}- \Delta_+^{\mu\alpha\beta}\left(\Box\delta^{\mu\nu} - \partial^\mu \partial^\nu
\right)\Delta_+^{\nu\alpha\beta} - v^\alpha\partial^\mu\partial^\nu
\Delta_+^{\mu\nu\alpha}\nonumber$$
$$\hspace{3cm}- \Delta_+^{\mu\nu\alpha}\partial^\mu \partial^\nu v^\alpha -
\Delta_-^{\mu\nu\rho} 
\partial^\mu\partial^\nu a^\rho - a^\rho\partial^\mu\partial^\nu \Delta_-
^{\mu\nu\rho}.\eqno(82)$$

The Lagrangian density for the matter fields is simpler to derive.  We consider separately the
kinetic and interaction parts, which upon using $\gamma^\alpha \gamma^\beta \gamma^\gamma
= \delta^{\alpha\beta} \gamma^\gamma - \delta^{\alpha\gamma}\gamma^\beta +
\delta^{\beta\gamma}\gamma^\alpha + \epsilon^{\alpha\beta\gamma\delta}
\gamma^5\gamma^\delta$ becomes
$$L_{matter}^{(2)} = - \sbody{i}{2} \Psi^{\dagger} \not{\!\partial} \Psi +
i\Psi^{\dagger\alpha\mu} (2
\gamma^\mu\partial^\nu - \delta^{\mu\nu} \not{\partial}) \Psi^{\alpha\nu}\nonumber$$
$$ - i \Psi^{\dagger\mu\nu} (\gamma^\nu \partial^\mu)\Psi - i
\Psi^\dagger
(\gamma^\nu\partial^\mu)\Psi^{\mu\nu}\eqno(83)$$
$$L_{matter}^{(3)} = - \sbody32 i \Psi^\dagger \gamma \cdot a \gamma_5 \Psi +
\Psi^{\alpha\mu{^{\dagger}}} \Sigma^{\alpha\mu} \gamma^\nu \Psi^{\nu\beta}v^\beta - v^\alpha
\Psi^{\alpha\mu{^{\dagger}}}\gamma^\mu \Sigma^{\beta\nu} \Psi^{\beta\nu}\nonumber$$
$$\;\;\;+ i \epsilon^{\alpha\mu\beta\rho} \Psi^{\alpha{\lambda^{\dagger}}} \gamma^\lambda
\gamma^\mu \gamma^\nu \Psi^{\nu\beta} a^\rho - i \Delta_+^{\mu\alpha\beta}
\Psi^{\alpha{\lambda^{\dagger}}} \gamma^\lambda \gamma^\mu \gamma^\nu
\Psi^{\nu\beta}\nonumber$$
$$\;\;\;\;\;+ 2i\, \Psi^\dagger \gamma^\mu \Psi^{\mu\nu}v^\nu + \sbody12 \Psi^\dagger \gamma
\cdot v
\Sigma^{\mu\nu} \Psi^{\mu\nu}  + \Psi^\dagger \Sigma^{\mu\nu} \gamma_5 \gamma^\alpha
\Psi^{\alpha\mu} a^\nu\nonumber$$
$$\;\;\;\;\;+ 2i\, v^\mu \Psi^{\alpha\mu{^{\dagger}}} \gamma^\nu \Psi + \sbody12
\Psi^{\mu\nu{^{\dagger}}}
\Sigma^{\mu\nu}  \gamma \cdot v \Psi - \Psi^{\alpha\mu{^{\dagger}}} \gamma^\mu
\Sigma^{\alpha\nu} \gamma_5 \Psi a^\nu\nonumber$$
$$+ \sbody{i}{2} \left(
\Psi^\dagger \gamma^\alpha \gamma^\mu \gamma^\nu \Psi^{\nu\beta}  - 
\Psi^{\alpha{\nu^{\dagger}}} \gamma^\nu \gamma^\mu \gamma^\beta\Psi \right)
\Delta_+^{\mu\alpha\beta}\; .\eqno(84)$$
It is curious to note that the only term in (84) which is bilinear in $\Psi$ and $\Psi^\dagger$
involves just the axial field $a_\mu$ and does not involve the vector field $v_\mu$ or the tensor
field $\Delta_+^{\mu\nu\lambda}$.

\section{Discussion}
In this paper we have examined a particular gauge invariant model in four dimensional Euclidean
space which has {\it{extended}} field transformations: we allow the gauge group indices on the
fields to participate in the Euclidean space $SO(4)$ field transformations.  This requires, of
necessity, that the gauge group in question has a four-dimensional representation.  In this paper
we have chosen $SO(4)$ itself as the gauge group, availing of the fundamental, or vector,
representation.  Gabrielli [1] availed of the gauge group $U(4)$.

The $SO(4)$ gauge model is defined by its action and the gauge transformations under which
the action is invariant.  The structure of the $SO(4)$ gauge model is very standard -- we choose
as action density the sum of a pure $SO(4)$ Yang-Mills $F^2$ term and a standard (for
Euclidean space) gauge-invariant matter-gauge field term $\Psi^{\dagger} (\not{\!\!p} +
\not{\!\!A})\Psi$ where the matter field is in the fundamental representation of the gauge group. 
However, the gauge potential $(A^\mu)^{\alpha\beta}$ and the matter field $\Psi^\mu$ are now
reducible under $SO(4)$.

In field theory models we always work with fields which transform irreducibly under the
symmetry transformations of the space the model is built on.  Consequently we investigated the
decomposition of the reducible gauge potential $t^{\mu\nu\lambda} = -i(A^\mu)^{\nu\lambda}$
and the reducible matter field $\Psi^\mu$ in terms of $SO(4)$ irreducible component fields.  The
results of this analysis can be presented in two ways which may be related by parity
transformations.

We consider first the decomposition of the gauge potential.  We found the $t^{\mu\nu\lambda}$
can be decomposed in terms of a vector field $v^\mu$, an axial vector field $a^\mu$, a rank $3$
tensor $\Delta_S^{\mu\nu\lambda} \left( = -\Delta_S^{\mu\lambda\nu} \right)$, self-dual with
respect to the $(\nu,\lambda)$ indices and traceless over the $(\mu, \nu)$ indices, and another
rank $3$ tensor $\Delta_A^{\mu\nu\lambda} \left( = -\Delta_A^{\mu\lambda\nu} \right)$, 
anti-self-dual with respect to the $(\nu , \lambda)$ indices and traceless over the $(\mu , \nu)$
indices.

This decomposition corresponds to
$$( \sbody12, \sbody12 ) \otimes [ (1,0) \oplus (0,1)] = (\sbody12, \sbody12 )_v \oplus (\sbody12,
\sbody12 )_a \oplus ( \sbody32 , \sbody12 )_S \oplus ( \sbody12 , \sbody32 )_A \;\; .\eqno(85)$$
Alternatively we can consider two linear combinations of $\Delta_S$ and $\Delta_A$ which are
dual to one another, $\Delta_{\pm} = \Delta_S \pm \Delta_A$, ${^*}\Delta_+ = \Delta_-$.

In treating the matter fields $\Psi^\mu$ we can choose to use $2$-component or $4$-component
spinor notation.  In the former case we found that $\Psi^\mu$ can be decomposed in terms of
$2$-component spinors in the $(\sbody12 , 0)$ and $(0, \sbody12 )$ representations, $\chi$ and
$\psi$ respectively, and $2$-component spinors in the $(\sbody12 , 1)$ and $(1, \sbody12 )$
representations, namely $\chi_A^{\alpha\beta} ( = - \chi_A^{\beta\alpha})$ and
$\psi_S^{\alpha\beta} ( = - \psi_S^{\beta\alpha})$ which are, respectively, anti-self-dual and 
self-dual.  In the $4$-component notation we have $\Psi$ in the $(\sbody12 , 0) \oplus(0,
\sbody12
)$ representation and $\Psi^{\alpha\beta} (= - \Psi^{\beta\alpha})$ in the $( \sbody12 , 1) \oplus
(1, \sbody12 )$ representation as the two components in the decomposition of $\Psi^\mu$.

Having the $SO(4)$-irreducible field structure of the model in hand, we proceeded to an
examination of the induced gauge transformations for each of these fields.  The resulting gauge
transformations, while complicated, have the distinctive feature of mixing the fields.  In a sense,
$(v^\mu, a^\mu, \Delta_S^{\mu\nu\lambda}, \Delta_A^{\mu\nu\lambda})$ can be thought of as
a multiplet of fields under the $SO(4)$ gauge transformations.  The same comment applies to
$(\chi , \chi_A^{\alpha\beta})$ and $(\psi , \psi_S^{\alpha\beta})$.

Finally, we considered the standard action density in terms of the irreducible field components.

Although Gabrielli [1] considered a different gauge group, and consequently the multiplet of
gauge fields in his case is different to the one we have found, our $t^{\mu\alpha\beta}$
corresponds to his $C_{\mu\alpha\beta}$.  In his paper the action density involves not only the
$L_{YM}$ which we have in eq. (79) but also the induced interactions with the other fields in
his multiplet $(\phi, \overline{\phi}, A, \overline{A}, T^S, \overline{T}^S, T^A,
\overline{T}^A)$ and their self-interaction and kinetic terms.  The explicit form of the action
density in terms of the fields contained within $C_{\mu\alpha\beta}$ was not considered.

It is interesting to note that we could have made a different choice of gauge group representation
for both the gauge potential and the matter field $\Psi$.  If instead of the fundamental
representation of $SO(4)$ we had chosen the $4$-component spinorial representation, the
resulting decomposition in terms of $SO(4)$ irreducible fields would be very different.  In that
case we would use $\sbody12 \Sigma^{\mu\nu}$ in place of $M^{\mu\nu}$ throughout.  The
gauge potential would correspond exactly to that considered by Gabrielli.  Furthermore the
extended $SO(4)$ field transformations would be
$$\hspace{-2cm}(A^\mu)^{ij}(x) \rightarrow (A^{\prime\mu})^{ij}(x^\prime) =
\Lambda^{\mu\nu}
\left[S(\lambda)A^\nu (x) S^{-1} (\lambda) \right]^{ij}\nonumber$$
$$ = \Lambda^{\mu\nu} S(\lambda)^{im}(A^\nu)^{mn}(x) S^{-1}
(\lambda)^{nj}\eqno(86)$$
$$\hspace{-2cm}\Psi_i^k(x) \rightarrow \Psi_i^{\prime k}(x^\prime) = S_{ij}(\lambda)
\left[S(\lambda)\Psi_j(x)\right]^k\nonumber$$
$$ = S_{ij}(\lambda) S^{km}(\lambda) \Psi_j^m(x).\eqno(87)$$
The decomposition of $\Psi_i^k$ into irreducible component fields inevitably includes fermionic
scalar and vector fields.  This type of result has previously been found in applying these ideas
in three Euclidean dimensions [7].

In an examination of the pure Yang-Mills sector of the model presented here, the presence of
instantons has previously been noted [2].

There are a number of directions in which the approach of this paper can be developed.  First
however, we should note a direction that is unlikely to be useful, namely, applying the analogous
approach in flat four dimensional Minkowski space: $SO(3,1)$ is not a compact group and, as
a consequence, there will be unwelcome ghost states [8].  A more promising direction of further
study would be to extend the gauge group from $SO(4)$ to $SO(4) \otimes G$ where $G$ could
be any compact Lie group.

The approach of gauge models with extended field transformations can also be applied in higher
dimensional Euclidean spaces.  For example, in five dimensions the gauge potential
$(A^\mu)^{\alpha\beta} = -(A^\mu)^{\beta\alpha}$ has 50 independent components.  The
decomposition of the gauge potential analogous to our discussion in Section 3 is
$$(A^\mu)^{\alpha\beta} = i T^{\mu\alpha\beta} = i \left[ \delta^{\mu\alpha} V^\beta -
\delta^{\mu\beta} V^\alpha + \epsilon^{\mu\alpha\beta\lambda\sigma} E^{\lambda\sigma} +
D^{\mu\alpha\beta} \right]\eqno(88)$$
where
$$D^{\mu\alpha\beta} = - D^{\mu\beta\alpha}\; ,\eqno(89a)$$
$$\hspace{-.5cm}D^{\mu\mu\beta} = 0 ,\eqno(89b)$$
$$\qquad E^{\lambda\sigma} = \epsilon^{\lambda\sigma\alpha\beta\gamma}
D^{\alpha\beta\gamma}\;
.\eqno(89c)$$
The vector field $V^\mu$ has 5 components, the anti-symmetric tensor $E^{\lambda\sigma}$ has
10 components while the rank 3 tensor field $D^{\mu\alpha\beta}$ has 45 components.  The
constraint (89c) leads to 50 independent components in total, effectively saying that the
$E^{\lambda\sigma}$ are not independent components.

It is of interest to consider the further decomposition of $V^\mu$ and $D^{\mu\alpha\beta}$ into
$SO(4)$ irreducible fields, thereby making contact with the analysis of Section 3.  It is
understood that the $SO(4)$ rotations are in the space spanned by $x^1, x^2, x^3, x^4$.  We also
identify the $SO(3)$ spin content where the $SO(3)$ rotations are in the space spanned by $x^1,
x^2, x^3$.  In what follows Greek indices are understood to take values 1 to 5, while Latin
indices take the values 1 to 4.  For the vector field we find
$$V^\mu = (V^i, V^5)\nonumber$$
where
$$V^i = ({\rm{spin}}\; 1, {\rm{spin}}\; 0)\nonumber$$
$$V^5 = ({\rm{spin}}\; 0).\nonumber$$
Next consider the fields $D^{\mu\alpha\beta}$ when none of the indices has the value 5, namely
$D^{ijk} = -D^{ikj}$.  This is just the gauge potential $t^{ijk}$ considered in detail in Section
3.  We identify
\begin{enumerate}
\item[(i)] the trace over the first pair of indices $v^k = \sbody13 D^{iik}$
$$v^k = ({\rm{spin}} \;1, {\rm{spin}} \;0)\nonumber$$
\item[(ii)] the trace over the first pair of indices of the dual $a^k = \sbody13\, {^*}D^{iik}$
$$a^k = ({\rm{spin}} \;1, {\rm{spin}} \;0)\nonumber$$
\item[(iii)] the self-dual and anti-self-dual parts of the traceless part of $D^{ijk}, \Delta^{ijk} =
D^{ijk} - \sbody13 \delta^{ij} D^{iik}$
$$\Delta_S^{ijk} = ({\rm{spin}} \;2, {\rm{spin}} \;1)\nonumber$$
$$\Delta_A^{ijk} = ({\rm{spin}} \;2, {\rm{spin}} \;1).\nonumber$$
\end{enumerate}
We need only consider one of the indices of $D^{\mu\alpha\beta}$ taking the value 5, as
$D^{\mu 55} = 0$ and $D^{55 k} = -D^{iik}$ due to the constraint (89b).  We first consider
$D^{ij 5}$ and note that this is traceless as (89b) implies $D^{\alpha\alpha 5} = D^{ii 5} = 0$.
We can then identify
\begin{enumerate}
\item[(iv)] the symmetric traceless part $T^{ij} (= T^{ji} ) = \sbody12 (D^{ij 5} + D^{ji
5})$
$$T^{ij} = ( {\rm{spin}} \;2, {\rm{spin}} \;1, {\rm{spin}} \;0)\nonumber$$
\item[(v)] the self-dual and anti-self-dual parts of the anti-symmetric part of $D^{ij 5}$, namely
$B_S^{ij}$, $B_A^{ij}$ where $B^{ij} = \sbody12( D^{ij 5} - D^{ji 5} )$.
$$B_S^{ij} = ({\rm{spin}} \;1)\nonumber$$
$$B_A^{ij} = ({\rm{spin}} \;1).\nonumber$$
\end{enumerate}
Finally, we consider $D^{5 ij} = -D^{5 ji} \equiv C^{ij}$
\begin{enumerate}
\item[(vi)] the self-dual and anti-self-dual parts $C_S^{ij}$, $C_A^{ij}$
$$C_S^{ij} = ({\rm{spin}} \;1)\nonumber$$
$$C_A^{ij} = ({\rm{spin}} \;1).\nonumber$$
\end{enumerate}
In all we count 3 spin 2 fields, 10 spin 1 fields and 5 spin 0 fields arriving at 50 independent
components as expected.  The mixing of the above fields under an $SO(5)$ Yang-Mills gauge
transformation is inevitable.  It is interesting to note that the
maximum spin we need to consider is spin two, in both four and five
dimensions.

The quantization of extended gauge models of the type proposed in this paper has not yet been
considered.  One of the first questions to be addressed in that procedure will be the most
appropriate form of gauge fixing.  It is important to identify which of the field components are
dynamical and which are gauge artifacts.  In this way it will be possible to identify clearly the
dynamical degrees of freedom.  Although it seems more efficient to do explicit calculations of
quantum effects in terms of the original reducible Yang-Mills and matter fields rather than in
terms of the $SO(4)$ irreducible fields, this will depend on the identification of dynamical
degrees of freedom.  We note in the action density many occurrences of
$\epsilon^{\mu\nu\alpha\beta}$ and $\gamma_5$; both quantities are intrinsically connected with
four
dimensions.  This indicates that the regularization scheme to be used should be one which
preserves the gauge invariance and does not alter the number of spatial dimensions.  Operator
regularization [9] is one such regularization scheme.

\section{Acknowledgements}

This work was supported by FORBAIRT under the 1998 International Collaboration Programme
and NSERC.  D.G.C. McKeon would like to thank NUI, Galway for their hospitality during his
stay there where this work was begun.  T.N. Sherry would like to thank UWO for their
hospitality
during his stay there where this work was completed.

\eject
\noindent{\Large\bf{Appendix A.  Spinor Notation in 4dE.}}\\
In this appendix we summarize the spinor notation in $4dE$ developed in
[3].\\
{\large\bf{A.1 $\gamma$-matrix conventions}}
$$\left\lbrace \gamma^\mu , \gamma^\nu \right\rbrace = 2 \delta^{\mu\nu} \qquad\qquad \mu ,
\nu = 1, \ldots, 4\nonumber$$
$$\gamma^{\mu\dagger} = \gamma^\mu \qquad\qquad\quad \mu = 1, \ldots, 4\nonumber$$
$$\gamma^\mu = \left[ \begin{array}{cc} 0 & \sigma^\mu \\
\overline{\sigma}^\mu & 0 \end{array}\right]\nonumber$$
$$\sigma^\mu = (i \vec{\tau}, 1), \quad \overline{\sigma}^\mu = (-i\vec{\tau}, 1) =
(\sigma^\mu)^\dagger\nonumber$$
$$\gamma_5 = \sbody{1}{4!} \epsilon^{\mu\nu\lambda\rho} \gamma^\mu \gamma^\nu
\gamma^\lambda \gamma^\rho = \left[ \begin{array}{cc} -1 & 0 \\ 0 & 1 \end{array} \right]\;,
\quad \epsilon^{1234} = 1\nonumber$$
$$\Sigma^{\mu\nu} = \sbody{i}{2} [\gamma^\mu , \gamma^\nu] = \Sigma^{\mu\nu\dagger} =
-
\sbody12 \epsilon^{\mu\nu\alpha\beta} \Sigma^{\alpha\beta} \gamma_5
 = \left[\begin{array}{cc} i\sigma^{\mu\nu} & 0 \\ 0 &
-i\dot{\sigma}^{\mu\nu}\end{array}
\right]\nonumber$$
$$\sigma^{\mu\nu} = \sbody14 (\sigma^\mu \overline{\sigma}^\nu - \sigma^\nu
\overline{\sigma}^\mu)\nonumber$$
$$\dot{\sigma}^{\mu\nu} = \sbody14 (\overline{\sigma}^\mu \sigma^\nu - \overline{\sigma}^\nu
\sigma^\mu)\nonumber$$
$$C = \left[ \begin{array}{cc} -i\tau_2 & 0 \\ 0 & i\tau_2 \end{array} \right] = -C^T = -
C^\dagger = -C^{-1} = C^*\nonumber$$
{\large\bf{A.2 2-spinors and 4-spinors}}
$$\Psi = \left[ \begin{array}{c} \psi_a \\ \chi^{\dot{a}} \end{array}\right]\qquad
\Psi^\dagger = \left[ \overline{\psi}^a,  -\overline{\chi}_{\dot{a}} \right]\nonumber$$
$$\Psi^C = \left[ \begin{array}{c} -\overline{\psi}_a \\ \overline{\chi}^{\dot{a}}
\end{array}\right]\qquad
\Psi^{C \dagger} = \left[\psi^a,  \chi_{\dot{a}} \right]\nonumber$$
Raising and lowering 2-spinor indices:
$$\psi_a = \epsilon_{ab} \psi^b\; , \;\;\psi^{b} = \epsilon^{ba} \psi_a\; , \;\;\chi_{\dot{a}} =
\epsilon_{\dot{a}\dot{b}} \chi^{\dot{b}}\; , \;\;\chi^{\dot{b}} =
\epsilon^{{\dot{b}}{\dot{a}}} \chi_{\dot{a}}\nonumber$$
$$\epsilon_{ab} = \left[ \begin{array}{cc} 0 & 1\\ -1 & 0\end{array}\right]_{ab}\qquad\qquad
\epsilon_{\dot{a}\dot{b}} = \left[ \begin{array}{cc} 0 & 1\\ -1 &
0\end{array}\right]_{\dot{a}\dot{b}}\nonumber$$
$$\epsilon_{ab}\epsilon^{bc} = \delta_a^{\;\;\;c}\qquad\qquad
\epsilon_{\dot{a}\dot{b}}\epsilon^{\dot{b}\dot{c}} = \delta_{\dot{a}}^{\;\;\;\dot{c}}\nonumber$$
Index summation:\\
undotted indices are summed top to bottom: $\psi^a \lambda_a = -\psi_a \lambda^a =
\lambda^a\psi_a$\\
dotted indices are summed bottom to top: $\chi_{\dot{a}} \phi^{\dot{a}} = -\chi^{\dot{a}}
\phi_{\dot{a}} = \phi_{\dot{a}}\chi^{\dot{a}}$\\
Complex conjugation of 2-spinors:\\
$*$ conjugation raises/lowers 2-spinor indices accompanied by a ($\pm$) ($+$ if the index is in
its ``natural position'', $-$ if not) and adds a conjugation bar above the spinor (or removes the
bar already there).  For an unconjugated spinor the natural position for either an undotted or
dotted index is as a lower index, so
$$(\psi_a)^* = \overline{\psi}^a \Longrightarrow (\overline{\psi}^a)^* = \psi_a\nonumber$$
$$(\chi_{\dot{a}})^* = \overline{\chi}^{\dot{a}} \Longrightarrow (\overline{\chi}^{\dot{a}})^*
= \chi_{\dot{a}}\nonumber$$
but
$$(\psi^a)^* = -\overline{\psi}_a \Longrightarrow (\overline{\psi}_a)^* = -\psi^a\nonumber$$
$$(\chi^{\dot{a}})^* = -\overline{\chi}_{\dot{a}} \Longrightarrow (\overline{\chi}_{\dot{a}})^*
= -\chi^{\dot{a}}\nonumber$$
and we deduce that for a conjugated spinor the natural position for an index is as an upper
index.\\
{\large\bf{A.3 Various properties of the $\sigma$-matrices:}}\\
Products of matrices
$\sigma^\mu_{\; a\dot{b}}, \;\; \overline{\sigma}^{\mu\; \dot{a}b}, \;\;
\sigma^{\mu\nu\;\;\;\;b}_{\;\;\;\;a},
\;\;\dot{\sigma}^{\mu\nu\dot{a}}_{\;\;\;\;\;\;\;\dot{b}}$
$$\sigma^\mu \overline{\sigma}^\nu = \delta^{\mu\nu} 1 + 2 \sigma^{\mu\nu}\nonumber$$
$$\overline{\sigma}^\mu \sigma^\nu = \delta^{\mu\nu} 1 + 2 \dot{\sigma}^{\mu\nu}\nonumber$$
$$\sigma^{\mu\nu}\sigma^{\kappa\lambda} = -\sbody14 (\delta^{\mu\kappa} \delta^{\nu\lambda}
- \delta^{\mu\lambda} \delta^{\nu\kappa}) 1 - \sbody14 \epsilon^{\mu\nu\kappa\lambda}
1\nonumber$$
$$\hspace{3.5cm} -\sbody12 (\delta^{\mu\kappa} \sigma^{\nu\lambda} + \delta^{\nu\lambda}
\sigma^{\mu\kappa} -\delta^{\mu\lambda} \sigma^{\nu\kappa} - \delta^{\nu\kappa}
\sigma^{\mu\lambda}) \nonumber$$
$$\dot{\sigma}^{\mu\nu} \dot{\sigma}^{\kappa\lambda} = -\sbody14 (\delta^{\mu\kappa}
\delta^{\nu\lambda} - \delta^{\mu\lambda} \delta^{\nu\kappa}) 1 + \sbody14
\epsilon^{\mu\nu\kappa\lambda} 1\nonumber$$
$$\hspace{3.5cm}  -\sbody12 (\delta^{\mu\kappa} \dot{\sigma}^{\nu\lambda} +
\delta^{\nu\lambda}
\dot{\sigma}^{\mu\kappa} -\delta^{\mu\lambda} \dot{\sigma}^{\nu\kappa} - \delta^{\nu\kappa}
\dot{\sigma}^{\mu\lambda}) \nonumber$$
$$\sigma^{\alpha\beta} \sigma^\gamma = -\sbody12 (\delta^{\alpha\gamma} \sigma^\beta -
\delta^{\beta\gamma}\sigma^\alpha) - \sbody12
\epsilon^{\alpha\beta\gamma\rho}\sigma^\rho\nonumber$$
$$\dot{\sigma}^{\alpha\beta} \overline{\sigma}^\gamma = -\sbody12 (\delta^{\alpha\gamma}
\overline{\sigma}^\beta - \delta^{\beta\gamma}\overline{\sigma}^\alpha) + \sbody12
\epsilon^{\alpha\beta\gamma\rho}\overline{\sigma}^\rho\nonumber$$
$$\sigma^\alpha \dot{\sigma}^{\beta\gamma} = \sbody12 (\delta^{\alpha\beta} \sigma^\gamma
-
\delta^{\alpha\gamma}\sigma^\beta) - \sbody12
\epsilon^{\alpha\beta\gamma\rho}\sigma^\rho\nonumber$$
$$\overline{\sigma}^\alpha \sigma^{\beta\gamma} = \sbody12 (\delta^{\alpha\beta}
\overline{\sigma}^\gamma -\delta^{\alpha\gamma}\overline{\sigma}^\beta) + \sbody12
\epsilon^{\alpha\beta\gamma\rho}\overline{\sigma}^\rho\nonumber$$
Fierz identities:
$$\delta_a^{\;\;b} \delta_f^{\;\;e} = \sbody12 (\delta_a^{\;\;e} \delta_f^{\;\;b} -
\sigma^{\mu\nu\;e}_{\;\;a} \sigma^{\mu\nu\;\;\;b}_{\;\;\;\;f} )\nonumber$$
$$\delta_a^{\;\;b} \delta^{\dot{c}}_{\;\;\dot{d}} = \sbody12 \sigma^{\mu}_{\;\;\;a\dot{d}}
\,\overline{\sigma}^{\mu\dot{c}b}\nonumber$$
Duality Relations
$${^*}\sigma^{\mu\nu} \equiv \sbody12 \epsilon^{\mu\nu\lambda\sigma}
\sigma^{\lambda\sigma} = \sigma^{\lambda\sigma}\nonumber$$
$${^*}\dot{\sigma}^{\mu\nu} = - \dot{\sigma}^{\mu\nu} \nonumber$$
Other identities
$$(\dot{\sigma}^{\mu\nu})^{\dot{a}}_{\;\;\;\dot{a}} = 0\nonumber$$
$$(\sigma^{\mu\nu})^a_{\;\;a} = 0\nonumber$$
$$(\sigma^\mu)^{a\dot{b}} = (\overline{\sigma}^\mu)^{\dot{b}a}\nonumber$$
$$(\sigma^\mu)_c^{\;\;{\dot{b}}} = (\overline{\sigma}^\mu)^{\dot{b}}_{\;c}\nonumber$$
$$(\sigma^\lambda \overline{\sigma}^\kappa)_{ba} = -(\sigma^\kappa
\overline{\sigma}^\lambda)_{ab}\nonumber$$
$$\sigma_{a\dot{c}}^\lambda \sigma^{\kappa\;\;\dot{c}}_{\;b} = (\sigma^\lambda
\overline{\sigma}^\kappa)_{ab} = -(\sigma^\kappa \overline{\sigma}^\lambda)_{ba}\nonumber$$
{\large\bf{A.4 Complex conjugation of multi-spinors}}

By a multi-spinor we mean any quantity which has more than one spinorial index (dotted or
undotted, upper or lower).  The action of $*$ on such an object is to
\begin{enumerate}
\item[(i)] raise a lower index,
\item[(ii)] lower an upper index
\end{enumerate}
in each case taking account of the natural position of the index,
\begin{enumerate}
\item[(iii)] invert the order of the indices
\item[(iv)] place a conjugation bar on the multispinor (or remove it from an already conjugated
multispinor).
\end{enumerate}
We can include the $\sigma$-matrices in this discussion by noting that if the multispinor is
Hermitian there is no need to introduce a bar.  Thus,
$$(V_{a\dot{b}})^* = \overline{V}^{\dot{b}a}\;\;, 
(\overline{V}^{\dot{b}a})^* = {V}_{a\dot{b}}\;\;, 
(V_a^{\;\dot{b}})^* = - \overline{V}_{\dot{b}}^{\;a}$$
$$(\sigma^\mu)_{a\dot{b}} {^*} = (\overline{\sigma}^\mu)^{\dot{b}a}\nonumber$$
$$(\sigma^{\mu\nu})_a^{\;b} {^*} = - (\sigma^{\mu\nu})_b^{\;a}\nonumber$$
$$(\sigma^{\mu\nu})^{ab*} = (\sigma^{\mu\nu})_{ab} = (\sigma^{\mu\nu})_{ba}\nonumber$$
$$({\dot{\sigma}}^{\mu\nu})^{\dot{a}}_{\;\;\;\dot{b}} {^*} = -
(\dot{\sigma}^{\mu\nu})^{\dot{b}}_{\;\;\;\dot{a}}\nonumber$$
$$(\dot{\sigma}^{\mu\nu})_{\dot{a}\dot{b}}\, {^*} = (\dot{\sigma}^{\mu\nu})^{\dot{b}\dot{a}}
= (\dot{\sigma}^{\mu\nu})^{\dot{a}\dot{b}}\nonumber$$
$$\epsilon_{ab} {^*} = \epsilon^{ba}\;\;, (\psi_{ab}^{\;\;\;\;\dot{c}})^* = -
\overline{\psi}_{\dot{c}}^{\;\;\;ba}\nonumber$$
$$(\psi_a^{\;\;b\dot{c}})^* = \overline{\psi}_{\dot{c}b}^{\;\;\;a}\;\;,
(\overline{\psi}_a^{\;\;b\dot{c}})^* =
-
\psi_{\dot{c}b}^{\;\;\;a},\;\;\;\;\;\;\; {\rm{etc.}}\nonumber$$
{\Large\bf{Appendix B.}}\\
{\large\bf{B.1 2-spinor indices for a 4-vector}}
$$V_{a\dot{b}} = V^\mu(\sigma^\mu)_{a\dot{b}} \Longrightarrow V^\mu = \sbody12
V_{a\dot{b}}(\overline{\sigma}^\mu)^{\dot{b}a}\nonumber$$
$$W^{\dot{a}b} = W^\mu(\overline{\sigma}^\mu)^{\dot{a}b} \Longrightarrow W^\mu =
\sbody12
W^{\dot{a}b}(\sigma^\mu)_{b\dot{a}}\nonumber$$
{\large\bf{B.2 2-spinor indices for self-dual part of rank-2 anti-symmetric
tensor}}
$$S^{\mu\nu} = - S^{\nu\mu}\;\;, {^*}S^{\mu\nu} = \sbody12\, \epsilon^{\mu\nu\alpha\beta}
S^{\alpha\beta} = S^{\mu\nu}\nonumber$$
$$S_a^{\;\;\;b} = S^{\mu\nu} (\sigma^{\mu\nu})_a^{\;\;\;b}\nonumber$$
$$\Longrightarrow S_a^{\;\;\;b}(\sigma^{\alpha\beta})_b^{\;\;\;a} = S^{\mu\nu}
Tr[\sigma^{\mu\nu}\sigma^{\alpha\beta}]\nonumber$$
$$\Longrightarrow S^{\alpha\beta} = -\sbody12\,
S_a^{\;\;\;b}(\sigma^{\alpha\beta})^b_{\;\;\;a}\nonumber$$
We note $S_{ab} = S_{ba}$.\\
{\large\bf{B.3 2-spinor indices for anti-self dual part of rank-2 anti-symmetric
tensor}}
$$A^{\mu\nu} = -A^{\nu\mu}\;\;, {^*}A^{\mu\nu} = \sbody12 \epsilon^{\mu\nu\alpha\beta}
A^{\alpha\beta} = -A^{\mu\nu}\nonumber$$
$$A^{\dot{a}}_{\;\;\;\dot{b}} =
A^{\mu\nu}({\dot{\sigma}}^{\mu\nu})^{\dot{a}}_{\;\;\;\dot{b}}\nonumber$$
$$\Longrightarrow A^{\dot{a}}_{\;\;\;\dot{b}}   
(\dot{\sigma}^{\alpha\beta})^{\dot{b}}_{\;\;\;\dot{a}} = A^{\mu\nu}
Tr[\dot{\sigma}^{\mu\nu} {\dot{\sigma}}^{\alpha\beta}]\nonumber$$
$$\Longrightarrow A^{\alpha\beta} = -\sbody12
A^{\dot{a}}_{\;\;\;\dot{b}}(\dot{\sigma}^{\alpha\beta})^{\dot{b}}_{\;\;\;\dot{a}}\nonumber$$
We note $A_{\dot{a}\dot{b}} = A_{\dot{b}\dot{a}}$.\\
{\large\bf{B.4 Special cyclic sum property of $\Delta_S$, $\Delta_A$, $\Delta_+$, and
$\Delta_-$}}
$$\Delta_i^{\mu\nu\lambda} + \Delta_i^{\nu\lambda\mu} + \Delta_i^{\lambda\mu\nu} = 0\;\;\;\;i
= S, A, +, -\nonumber$$
where $\Delta_i^{\mu\nu\lambda} = -\Delta_i^{\mu\lambda\nu}$ and $\Delta_i^{\mu\mu\lambda}
= 0$.  To see this we define
$$H_i^{\mu\nu\lambda} = \Delta_i^{\mu\nu\lambda} + \Delta_i^{\nu\lambda\mu} +
\Delta_i^{\lambda\mu\nu}\nonumber$$
and consider the dual of $H_i^{\mu\nu\lambda}$ with respect to $(\nu\lambda)$ for $i = S,A$. 
We find
$${^*}H_i^{\mu\nu\lambda} = \sbody12 \epsilon^{\nu\lambda\alpha\beta} \left[
\Delta_i^{\mu\alpha\beta} + \Delta_i^{\alpha\beta\mu} +
\Delta_i^{\beta\mu\alpha}\right]\nonumber$$
$$= \pm \left[\Delta_i^{\mu\alpha\beta} \mp \epsilon^{\nu\lambda\alpha\beta}
\Delta_i^{\alpha\beta\mu} \right]\nonumber$$
$$= \pm \left[\Delta_i^{\mu\alpha\beta} +\sbody12 \epsilon^{\nu\lambda\alpha\beta} 
\epsilon^{\beta\mu\rho\tau} \Delta_i^{\alpha\rho\tau} \right]\nonumber$$
$$= \pm \left[\Delta_i^{\mu\alpha\beta} -  \Delta_i^{\mu\alpha\beta} \right]\nonumber$$
$$= 0\nonumber$$
from which the result follows.\\
{\large\bf{B.5 Special property of anti-symmetric tensors}}\\
If $A^{\mu\nu} = -A^{\nu\mu}$ and $B^{\mu\nu} = -B^{\nu\mu}$ and the anti-symmetric
tensors $T$ and $X$ are defined by
$$T^{\alpha\beta} = A^{\mu\alpha}B^{\mu\beta} - A^{\mu\beta}B^{\mu\alpha}\nonumber$$
$$X^{\alpha\beta} = {^*}A^{\mu\alpha}B^{\mu\beta} -
{^*}A^{\mu\beta}B^{\mu\alpha}\nonumber$$
then $X$ is the dual of $T$,
$${^*}X^{\alpha\beta} = T^{\alpha\beta}.\nonumber$$
\end{document}